\documentstyle[12pt]{article}
\newcommand{\newsection}{    
\setcounter{equation}{0}
\section}
\newcommand{\tr}[1]{\,{\rm tr}\,#1}
\newcommand{\ntr}[1]{\,\frac {\rm tr}{N}\,#1}
\def\e{{\,\rm e}\,}

\def\eop{\vspace*{\fill}\pagebreak}
\def\be{\begin{equation}}
\def\ee{\end{equation}}
\def\bea{\begin{eqnarray}}
\def\eea{\end{eqnarray}}
\def\LA{\left\langle}
\def\RA{\right\rangle}
\newcommand{\Di}{\,\hbox{Disc}_\nu\,}

\newcommand{\rf}[1]{(\ref{#1})}
\newcommand{\eq}[1]{Eq.~(\ref{#1})}

\def\d{\partial}
\def\L{\Lambda}
\def\l{\lambda}
\def\om{\omega}
\newcommand{\ie}{{\it i.e.}\ }
\newcommand{\p}{{\prime}}
\newcommand{\ra}{\rightarrow}
\newcommand{\fr}[2]{{\textstyle {#1 \over #2}}}
\newcommand{\ci}{\int_{C_1}\frac{d\omega}{2\pi i}}
\newcommand{\cione}{\int_{C_1}\frac{d\omega}{2\pi i}\;\frac{
V^\prime(\omega)}{\lambda-\omega}}

\newcommand{\eps}{\varepsilon}
\newcommand{\non}{\nonumber \\*}
\newcommand{\VVp}{{\cal V}^\prime}
\newcommand{\re}{\,\hbox{Re}\,}
\newcommand{\im}{\,\hbox{Im}\,}

\begin{document}

\begin{titlepage}
\begin{flushright}
ITEP-YM-6-93 \\
September, 1993 \\
{\small hep-th/9310191}
\end{flushright}
\vspace{1cm}

\begin{center}
{\LARGE  Matrix Models of 2$D$ Gravity and Induced QCD}
\end{center} \vspace{1cm}
\begin{center} {\large Yu.\ Makeenko}
\footnote{E--mail:  \ makeenko@vxitep.itep.msk.su \ /
\ makeenko@nbivax.nbi.dk \ / \
makeenko@vxdesy.desy.de \ }
\\ \mbox{} \\
{\it Institute of Theoretical and Experimental Physics, 117259 Moscow, RF}
\\ \vskip .2 cm and  \\  \vskip .2 cm
{\it The Niels Bohr Institute, 2100 Copenhagen, DK}

\end{center}

\vskip 1.5 cm
\begin{abstract}
I review some recent works on the Hermitean one-matrix and $d$-dimensional
gauge-invariant matrix models. Special attention is paid to solving the
models at large-$N$ by the loop equations. For the one-matrix model the
main result concerns calculations of higher genera, while for the
$d$-dimensional model the large-$N$ solution for a logarithmic potential is
described. Some results on fermionic matrix models are briefly reviewed.
\end{abstract}

\vspace{2.5cm}
\noindent
Talk at the Workshop on
{\sl Quantum Field Theoretical Aspects of High Energy Physics},
Kyffhaeuser, Germany, September 20--24, 1993

\eop
\end{titlepage}
\setcounter{page}{2}

\newsection{Introduction}

Matrix models are usually associated~\cite{Kaz85} with discretized random
surfaces (or string theory) in a $d\leq1$-dimensional embedding space and, in
particular, with $2D$ quantum gravity.
The simplest Hermitean one-matrix model corresponds~\cite{Kaz89,ds}
to pure $2D$ gravity while a chain of Hermitean matrices
describes \mbox{\cite{2,3,4}} $2D$ gravity interacting with $d\leq1$ matter.
A natural multi-dimensional extension of this construction is
associated~\cite{KM92} with induced lattice gauge theories. The matrix models
which describe induced QCD can be constructed in a similar way~\cite{KhM92b}.

\subsection{Random surfaces}

The typical problems which reduce to matrix models are associated with
a statistical ensemble of random surfaces whose partition function is defined
generically as
\be
Z_{RS} = \sum_S \e^{-\sigma A(S)} \,.
\label{Zrs}
\ee
Here $A(S)$ is the area of the surface $S$ which can be either closed or open
and $\sigma$ stands for the string tension.

There is a lot of examples of such systems in quantum field theory:
\begin{itemize}
\item \vspace{-8pt}
Strings either fundamental (graviton) or secondary (hadrons).
\item \vspace{-8pt}
$3D$ Ising model (the boundary between different phases is two-dimensional).
\item \vspace{-8pt}
Lattice gauge theory at strong coupling.
\item \vspace{-8pt}
$1/N$-expansion of QCD~\cite{Hoo74}.
\item \vspace{-8pt}
$2D$ quantum gravity.
\end{itemize}
\vspace{-7pt}

The last system is described by the Euclidean partition function
\be
Z_{2D}=\int Dg \e^{-\int d^2x \sqrt{g} \left( \L-\frac{1}{4\pi G} R \right)} =
\int Dg \e^{-\int d^2x \sqrt{g} \L + \frac{\chi}{G} }
\label{2Dgravity}
\ee
where $\L$ stands for the cosmological constant and $\chi$ is the
Euler characteristics of the $2D$ world. The path integral in \eq{2Dgravity}
is over all metrics $g_{\mu\nu}(x)$.

\subsection{Dynamical triangulation \label{Dt}}

The idea of dynamical triangulation of random surfaces is to approximate the
surface by a set of equilateral triangles. The coordination number (the number
of triangles meeting at a vertex) is not necessarily equal to six which is
associated with internal curvature of the surface.  The partition
function~\rf{2Dgravity} is approximated by~\cite{Kaz85}
\be
Z_{DT} = \sum_g
\e^{\frac 2G (1-g)} \sum_{T_g} \e^{-\L n_t} \label{Zdt}
\ee
where one splits the sum over all the triangles into the sum over the genera,
$g$, and the sum over all possible triangulations, $T_g$,  at fixed genus $g$.
Remember that $g=0$ for a sphere, $g=1$ for a torus and $\chi = 2(1-g)$.
In~\rf{Zdt}  $n_t$
stands for the number of triangles which is not fixed and is a
dynamical variable.

The exponential suppression with $n_t$ provides the convergence of the sum over
$T_g$ in~\rf{Zdt} at least for large enough $\L$. However, the sum can diverge
for some values
of $\L$ due to the entropy factor (the number of graphs).
It is crucial for what follows that the total number of graphs of genus $g$
with $n$ triangles grows at large $n$ as a power of $n$~\cite{KNN77}:
\be
\sum_{T_g} \delta(n_t-n) =\e^{\L_c n} n^{-b_g} \left(1+ {\cal O}(n^{-1})\right)
\,,
\ee
where $\L_c$ does not depend on $g$. For this reason the genus $g$
contributions to the string susceptibility \be f= \frac{\d^2}{\d \L^2} Z_{DT}
\sim \sum_g \e^{\frac 2G (1-g)} (\L-\L_c)^{-\gamma_g}\,,~~~~\gamma_g=-b_g +3
\label{sucseptibility}
\ee
simultaneously diverge as $\L\ra\L_c+0$. This is the point where the continuum
limit is reached and the discrete partition function $Z_{DT}$ approaches the
continuum one $Z_{2D}$.

A similar dynamical triangulation can be written~\cite{Kaz85} for the partition
function~\rf{Zrs} of $2D$ surfaces embedded in a $d$-dimensional space.  The
case of $2D$ gravity is associated with $d=0$.

\subsection{Large-$N$ matrix models}

The partition function~\rf{Zdt} can be represented as a matrix model. The dual
graph for the set of triangles coincides with the graph in a $d=0$ quantum
field
theory with a cubic interaction as is depicted in Fig.~\ref{dual}.
\begin{figure}[tbp]
\unitlength=0.50mm
\begin{picture}(44.0,100.00)(-95,40)
\thicklines
\put(40.00,120.00){\line(1,0){40.00}}
\put(80.00,120.00){\line(3,-5){20.00}}
\put(100.00,86.67){\line(-3,-5){20.00}}
\put(80.00,53.33){\line(-1,0){40.00}}
\put(40.00,53.33){\line(-3,5){20.00}}
\put(20.00,86.67){\line(3,5){20.00}}
\put(40.00,120.00){\line(3,-5){20.00}}
\put(60.00,86.67){\line(3,5){20.00}}
\put(80.00,53.33){\line(-3,5){20.00}}
\put(60.00,86.67){\line(1,0){40.00}}
\put(20.00,86.67){\line(1,0){40.00}}
\put(60.00,86.67){\line(-3,-5){20.00}}
\thinlines
\put(60.00,108.50){\line(5,-3){20.00}}
\put(80.00,96.50){\line(0,-1){20.00}}
\put(80.00,76.50){\line(-5,-3){20.00}}
\put(60.00,64.50){\line(-5,3){20.00}}
\put(40.00,76.50){\line(0,1){20.00}}
\put(40.00,96.50){\line(5,3){20.00}}
\put(60.00,108.50){\line(0,1){24.00}}
\put(60.00,64.50){\line(0,-1){24.00}}
\put(80.00,96.50){\line(5,3){20.00}}
\put(80.00,76.50){\line(5,-3){20.0}}
\put(40.00,76.50){\line(-5,-3){20.00}}
\put(40.00,96.50){\line(-5,3){20.00}}
\end{picture}
\caption[x]   {\hspace{0.2cm}\parbox[t]{13cm}
{\small
   A graph constructed from equilateral triangles
   (depicted in the bold lines) and
   its dual one (depicted in the thin lines).  }}
   \label{dual} \end{figure}
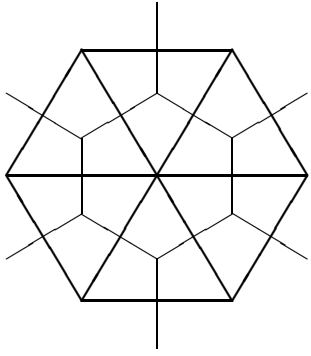 The
precise statement is that $Z_{DT}$ equals to the partition function of the
$N\times N$ Hermitean one-matrix model
\be Z_{1M} \equiv \e^{N^2 F}=\int d\Phi
\e^{-N\tr V( \Phi)} \label{1mamo} \ee
with $N=\exp{(1/G)}$ and the cubic
coupling constant $t_3=\exp{(-\L_c)}$.  The integration measure in \eq{1mamo}
is \be
d \Phi = \prod_{i>j}^N d \re \Phi_{ij} \, d \im \Phi_{ij}
\prod_{i=1}^N d \Phi_{ii}
\ee
and $V(\Phi)= \fr 12 \Phi^2 + t_3 \Phi^3$ is a cubic potential.

The fact that $Z_{DT}=Z_{1M}$ can be proven analyzing
the ``fat-graph'' expansion
of~\rf{1mamo} with the propagator
\be
(2\pi)^{-\frac{N^2}{2}}\int d\Phi \e^{-\frac N2 \tr ( \Phi^2)} \Phi_{ij}
\Phi_{kl} = \frac 1N \delta_{il} \delta_{kj}
\ee
which leads for $\;\log{Z_{1M}}\;$
to the factor $N^{2-2g}$ associated~\cite{Hoo74} with a graph of genus $g$.

The partition function~\rf{1mamo} with the general potential
\be
V(\Phi)=\sum_{j=0}^{\infty} t_j \, \Phi^j
\label{potential}
\ee
is associated with a discretization by regular polygons with $j\geq3$ vertices
whose area is $j$--$2$ times the area of the equilateral triangle.

\subsection{Double-scaling limit}

A question arises how the system described by the partition function~\rf{1mamo}
can undergo a phase transition for $\L\ra\L_c$ which is associated, as is
discussed in Subsect.~\ref{Dt}, with the continuum limit.  While the system is
at $d=0$, a (third-order) phase transition of the Gross--Witten
type~\cite{GW80} is possible as $N\ra\infty$ when the number of degrees of
freedom becomes infinite. Therefore, the continuum limit is reached at
$N\ra\infty$ and $\L\ra\L_c$.

The $N=\infty$ limit corresponds to planar diagrams or genus zero (the
spherical approximation).  Higher genera are suppressed as $N^{-2g}$.

One can utilize, however, the fact that $\gamma_g$, which is defined by
\eq{sucseptibility}, linearly depends on $g$~\cite{KM87}:
\be
\gamma_g= 2 + \frac 52 (g-1)\,.
\label{gamma}
\ee
Therefore, the parameter of the genus expansion near the critical point is \be
{\cal G} =\frac{1}{N^2(\L-\L_c)^\frac 52}
\ee
and can be made finite if $ (\L-\L_c) \sim N^{-\frac 45} $ as
$N\ra\infty$~\cite{ds}.  This special limit when the couplings  reach critical
values in a $N$-dependent way as $N\ra\infty$ is called the double scaling
limit. The double scaling limit of the Hermitean one-matrix
model allowed~\cite{ds} to construct the genus expansion of $2D$
quantum gravity.

\subsection{The Kazakov--Migdal model and induced QCD}

A natural $d>1$-dimensional extension of~\rf{1mamo}
is the Kazakov--Migdal model~\cite{KM92} which
is defined by the partition function
\be
Z_{KM}=\int \prod_{x,\mu} dU_{\mu}(x) \prod_x
 d\Phi_x \e^{ \sum_x N \tr{}\left(-V(\Phi_x)+
\sum_{\mu=1}^D\Phi_x U_\mu(x)\Phi_{x+\mu}U_\mu^\dagger(x)\right)}\,.
\label{spartition}
\ee
Here the integration over the gauge field $U_\mu(x)$ is over the Haar measure
on $SU(N)$ at each link of a $d$-dimensional lattice with $x$ labeling
its sites.
The model~\rf{spartition}
obviously recovers the standard $d\leq1$ matrix chain if the lattice is just
a one-dimensional sequence of points for which the gauge field can be absorbed
by a unitary transformation of $\Phi_x$.

The large-$N$ solution of the
Kazakov--Migdal model in the strong coupling phase
\mbox{\cite{Mig92a}--\cite{Mak92}}
is associated~\cite{KSW92} with an unbroken extra $Z(N)$ symmetry
of the partition function~\rf{spartition}  and,
therefore, with infinite string tension (see~\cite{Mak92b} for a review).
One can easily modify the model in order to have a phase
transition (with decreasing the bare mass parameter) after which the $Z(N)$
symmetry is broken in some sense and normal area law associated with finite
string tension is restored. While the arguments~\cite{KhM92b} are based on
the mean field analysis, they look quite reasonable because the phase
transition occurs only for systems which are not asymptotically free.

The simplest model of this type is the adjoint fermion model
which is defined by the partition function~\cite{KhM92b}
\be
Z_{AFM}=\int \prod_{x,\mu} dU_{\mu}(x) \prod_x  d\Psi_x
d\bar{\Psi}_x \e^{-S_{F}[\Psi,\bar{\Psi},U]}
\label{fpartition}
\ee
where $\Psi_x$ and $\bar{\Psi}_x$ are the $N\times N$ matrices whose
elements are independent anticommuting Grassmann variables,
$S_{F}[\Psi,\bar{\Psi},U]$ is the lattice fermion action
\be
  S_{F} =
 \sum_x N \tr{\Big(V_F({\bar{\Psi}_x \Psi_x}) -
\sum_{\mu=1}^D [\bar{\Psi}_x P_\mu^-U_\mu(x)\Psi_{x+\mu}U_\mu^\dagger(x)}
+\bar{\Psi}_{x+\mu} P_\mu^+U_\mu^\dagger(x)\Psi_x U_\mu(x)]\Big)
\label{faction}
\ee
and $P_\mu^\pm$ are the standard projectors.

\newsection{Higher genera in one-matrix model}

The Hermitean one-matrix model was first solved in genus zero in
Ref.~\cite{BIPZ78} by the method of the saddle-point integral equations
for the spectral density. The more powerful
orthogonal polynomial technique allowed to
calculate the partition function up to genus two for the quartic
interaction~\cite{Bes79}. I review in this section the method of solving
the Hermitean one-matrix model which is based on the loop equations
and provides an algorithm for genus by genus calculations.
The explicit results~\cite{ACM92,ACKM93} on calculation of the partition
function with an arbitrary potential and all correlators are
presented up to genus two.

\subsection{Loop equation}

All correlators of the Hermitean one-matrix model~\rf{1mamo} can be
obtained from the Laplace image of the Wilson loop \be W(\l)= \LA \ntr
\frac{1}{\l-\Phi} \RA \label{defW} \ee where the averaging is w.r.t.\ the same
measure as in~\rf{1mamo}.  As is explained in Ref.~\cite{Kaz89}, $W(\l)$
is associated with the sum over discretized open surfaces with one
fixed boundary.

The correlator~\rf{defW} can be obtained from the free energy, $F$,
by applying the loop insertion operator,
$d/{dV(\l)}$:
\be
W(\l)=\frac{dF}{dV(\l)}\,,~~~~~~~
\frac{d}{dV(\l)}\equiv
-\sum_{j=0}^{\infty}\frac{1}{\l^{j+1}}\frac{\d}{\d t_{j}}\;.
\label{loopaverage}
\ee

$W(\l)$ is determined by the loop equation (see \cite{Mig83,Mak91}
for a review)
\be
\cione W(\om) = W^2(\l) +\frac{1}{N^2} \frac{d}{dV(\l)} W(\l)
\label{leq}
\ee
where the contour $C_1$ encloses counterclockwise
singularities of $W(\omega)$.
The contour integration acts as a projector
picking up negative powers of $\l$.
The second term on the r.h.s.\
of the loop equation~\rf{leq} is expressed via $W(\l)$, so that
\eq{leq} is closed and unambiguously determines $W(\l)$ imposing the boundary
condition $\l W(\l)\ra 1$ as $\l\ra \infty$.

The genus expansion of $W(\l)$ and of the free energy, $F$, is defined by
\be
W(\l)=\sum_{g=0}^{\infty}\frac{1}{N^{2g}}\;
W_{g}(\l)\,,~~~~~
F=\sum_{g=0}^{\infty}\frac{1}{N^{2g}}F_g~~~~~\hbox{with}~~~
W_g(\l) = \frac{d F_g}{dV(\l)} \,.
\label{genus}
\ee

\subsection{Genus zero solution}

To leading order in $1/N^2$ one can disregard  the second term
on the r.h.s.\ of \eq{leq} which reduces to the quadratic equation
\be
\cione W(\om) = W^2(\l) \,.
\label{lleq}
\ee

The one-cut solution to \eq{lleq} reads~\cite{Mig83}
\be
W_{0}(\l)= \int_{C_1} \frac{d \om}{4\pi i} \frac{V^\p(\om)}{\l-\om}
\sqrt{\frac{(\l-x)(\l-y)}{(\om-x)(\om-y)}}
\label{one-cut}
\ee
where $x$ and $y$ are determined by
\be
\ci\; \frac{V^\p(\om)}{\sqrt{(\om-x)(\om-y)}}=0\,,
\hspace{1.0cm}
\ci\; \frac{\om V^\p(\om)}{\sqrt{(\om-x)(\om-y)}}=2\,.
 \label{xandy}
\ee
The formulas~\rf{one-cut}, \rf{xandy} solves \eq{lleq} when $V^\p(\om)$ is
polynomial (which is usually associated with the one-matrix model)
or has singularities outside the cut $[y,x]$.

Doing the contour integral in~\rf{one-cut} by
taking the residues at $\om=\l$ and
$\om=\infty$, one finds
\be
W_0(\l) =\frac{1}{2}\left\{V^\p(\l)-M(\l)\sqrt{(\l-x)(\l-y)}\right\}
\label{zeroint}
\ee
where $M(\l)$ is a polynomial in $\l$ of degree $J$--$2$ if $V(\l)$
is that of degree $J$.
This form of the genus zero solution is convenient to determine
the spectral density, $\rho(\l)$, which describes the distribution of
eigenvalues of the matrix $\Phi$ at the large-$N$ saddle point.
Calculating the discontinuity of $W(\l)$ across the cut $[y,x]$, one gets
\be
\rho(\lambda) \equiv \im W(\l)= \frac{1}{\pi} M(\lambda)
\sqrt{(\lambda-y)(x-\lambda)} \hspace{0.7cm}  \lambda\in [y , x]
\,.
\label{evden} \ee
\eq{xandy} guarantees that
$\int d\l \,\rho (\lambda)=1$.

The spectral density $\rho(\l)$ given by~\rf{evden}
vanishes under normal circumstances as a square root
at both ends of its support. The critical behavior emerges when some of the
roots of $M(\lambda)$ approach the end point $x$ or $y$.

\subsection{The iterative procedure}

The iterative procedure of solving the loop equation is based on the genus
zero solution~\rf{one-cut}.
Inserting the genus expansion~\rf{genus} in~\eq{leq}, one gets
the following equation for $W_g(\l)$ at $g\geq1$:
\be
\cione W_g(\om)
-2\,W_{0}(\l)\, W_g(\l) = \sum_{g'=1}^{g-1}
W_{g'}(\l)\;W_{g-g'}(\l)
+\frac{d}{d V(\l)}W_{g-1}(\l) \,,
\label{loopg}
\ee
which expresses $W_g(\l)$ entirely in terms of
$W_{g^\p}(\l)$ with $g^\p<g$. This makes it possible to
solve \eq{loopg} iteratively genus by genus.

The iterative procedure simplifies if one introduces, instead of the coupling
constants
$t_j$, the moments $M_k$ and $J_k$ defined for $k\geq 1$ by \be M_{k}=\ci\,
\frac{V^\p(\om)}{(\om-x)^{k+1/2}\,(\om-y)^{1/2}} \,,~~~~~ J_{k}=\ci\,
\frac{V^\p(\om)}{(\om-x)^{1/2}\,(\om-y)^{k+1/2}} \,.  \label{moments} \ee These
moments depend on the coupling constants $t_j$'s both explicitly and via $x$
and $y$ which are determined by Eq.~\rf{xandy}.
Notice that $M_k$ and $J_k$ depend explicitly only on $t_j$ with $j\geq k+1$.

The main motivation for
introducing the moments~\rf{moments}
 is that $W_g(\l)$ depends only on $2\times (3g-1)$
lower moments ( $2\times (3g-2)$ for $F_g$ )~\cite{ACM92,ACKM93}.
This is in contrast to the $t$-dependence
of $W_g$ and $F_g$ which always depend on the infinite set
of $t_j$'s ($1\leq j <\infty$).

\subsection{Genus one and two results}

To find $F_g$, one first solves \eq{loopg} for $W_g(\l)$ and then uses
the last equation in \rf{genus}. The result in genus one
reads~\cite{ACM92}
\be
F_1=-\frac{1}{24}\ln M_1 -\frac{1}{24}\ln J_1 -\frac{1}{6} \ln d
\label{hermF1}
\ee
where $d=x-y$.

An analogous calculation in genus two yields~\cite{ACKM93}
\bea
&F_2=
  -{{119}\over {7680\,{{J_1}^2}\,{{d }^4}}} -
   {{119}\over {7680\,{{M_1}^2}\,{{d }^4}}}
 +    {{181\,J_2}\over
     {480\,{{J_1}^3}\,{{d }^3}}}
-    {{181\,M_2}\over
     {480\,{{M_1}^3}\,{{d }^3}}}
+ {{3\,J_2}\over
     {64\,{{J_1}^2}\,M_1\,{{d }^3}}}
 -    {{3\,M_2}\over
     {64\,J_1\,{{M_1}^2}\,{{d }^3}}}
     & \nonumber \\* &
-  {{11\,{{J_2}^2}}\over
     {40\,{{J_1}^4}\,{{d }^2}}}
-{{11\,{{M_2}^2}}\over
     {40\,{{M_1}^4}\,{{d }^2}}}
     + {{43\,M_3}\over
     {192\,{{M_1}^3}\,{{d }^2}}}
+   {{43\,J_3}\over
     {192\,{{J_1}^3}\,{{d }^2}}} +
   {{J_2\,M_2}\over
     {64\,{{J_1}^2}\,{{M_1}^2}\,{{d }^2}}}
-    {{17}\over {128\,J_1\,M_1\,{{d }^4}}}  &\nonumber \\* &
+   {{21\,{{J_2}^3}}\over
     {160\,{{J_1}^5}\,d }}
-   {{29\,J_2\,J_3}\over
     {128\,{{J_1}^4}\,d }} +
   {{35\,J_4}\over {384\,{{J_1}^3}\,d }} -
   {{21\,{{M_2}^3}}\over
     {160\,{{M_1}^5}\,d }} \
+ {{29\,M_2\,M_3}\over
     {128\,{{M_1}^4}\,d }} -
   {{35\,M_4}\over {384\,{{M_1}^3}\,d }}\,. &
\label{f2complete}
\eea
Some of these coefficients have an interpretation in terms of the intersection
indices on moduli space~\cite{Wit90}, the others are associated with
characteristics of the discretized moduli space~\cite{Che93}.

Since $F_g$ is known, the genus $g$ contribution to any connected correlator
\be
\LA \ntr \left(\Phi^{i_1}\right) \ldots \ntr \left(\Phi^{i_s}\right)  \RA_{g}
= N^{2-2s}\frac{\d}{\d t_{i_1}} \ldots \frac{\d}{\d t_{i_s}} F_g
\ee
can be calculated by the differentiation.
It depends on at most $2\times (3g-2+s)$
lower moments~\cite{ACM92,ACKM93}.

To obtain explicit formulas, say, for the symmetric quartic potential when all
$t_j=0$ except $t_2$ and $t_4$, one should solve \eq{xandy} for $x=-y$:  \be
x^2=-\frac{t_2}{3t_4}+\sqrt{\left(\frac{t_2}{3t_4}\right)^2+\frac{4}{3t_4}}\,,
\ee
and express the moments~\rf{moments} via $t_2$ and $t_4$
 which is given by
algebraic formulas
\be
M_1=J_1=2t_2+6t_4x^2\,,~~~M_2=-J_2=8t_4x \,,~~~M_3=J_3=4t_4 \,,
{}~~~M_k=J_k=0~~~~\hbox{for}~~ k\geq 4 \,.
\ee
The results for $F_1$ and $F_2$ in the case of the quartic potential
are in agreement with those of Ref.~\cite{Bes79}.

\newsection{Solving matrix models at $d>1$}

The Kazakov--Migdal model was originally studied at large-$N$ by the
Riemann-Hilbert method~\cite{Mig92a}. The explicit solution for the
quadratic potential~\cite{Gro92} was reproduced~\cite{Mak92} by the
loop equations. The equivalence of the two methods was shown
for an arbitrary potential in
Ref.~\cite{DMS93} where the relation to the Hermitean two-matrix model
was utilized. This approach allowed to solve explicitly the Kazakov--Migdal
model with a logarithmic potential~\cite{Mak93} and the adjoint fermion model
with the quadratic potential~\cite{MZ93} in the strong coupling phase.

\subsection{Loop equation for one-link correlator}

Let us define for the Kazakov--Migdal model~\rf{spartition} the loop
average and the one-link correlator, respectively, by
\be
W(\l)= \LA\ntr{}\Big(\frac{1}{\l-\Phi_x}\Big)\RA \,,~~~~~
G(\nu,\lambda) = \LA\ntr{}\Big(
\frac{1}{\nu-\Phi_x}U_{\mu}(x)\frac{1}{\lambda-\Phi_{x+\mu}}
U_\mu^\dagger(x) \Big) \RA \,.
\label{defG}
\ee
The definition of $W(\l)$ is similar to \eq{defW} while
$G(\nu,\l)$, which is symmetric in $\nu$ and $\l$
due to invariance of the Haar measure, $dU$, under
the transformation $U\ra U^\dagger$, is absent in the one-matrix model.
Expanding $G(\nu,\l)$ in $1/\nu$, one gets
\be
G(\nu,\l)=\frac{W(\l)}{\nu} + \sum_{n=1}^\infty
\frac{G_n(\l)}{\nu^{n+1}} ~~,~~~~~~~
G_n(\l)=\LA\ntr{}\Big( \Phi^n_x U_\mu(x)\frac{1}{\l-\Phi_{x+\mu}}
U^\dagger_\mu(x) \Big)\RA \,.
\label{bcG}
\ee

The correlator $G(\nu,\l)$ obeys in the large-$N$
limit the following equation~\cite{DMS93}
\be
 \int_{C_1} \frac{d \om}{2\pi i}
\frac{\VVp(\om)}{\nu - \om}\,G(\om, \l)=
W(\nu)\, G(\nu, \l) + \l G(\nu, \l) - W(\nu) \, ,
\label{main}
\ee
where the contour $C_1$ encircles counterclockwise the cut (or cuts)
of the function $G(\om,\l)$
and
\be
\VVp(\om)\equiv V^\prime(\om)-(2d-1) F(\om)\,.
\label{defL}
\ee

The function
\be
F(\om)=\sum_{n=0}^\infty F_n \om^n\,,~~~~~~
F_0=\ntr{} \Big( \Phi - \sum_{n=1}^\infty F_n \Phi^n \Big)
\label{defF_0}
\ee
is determined by the pair correlator of the gauge fields
\be
\frac{\int dU\,\e^{N\tr{}( \Phi U
\Psi U^\dagger)} \ntr{} \Big(t^aU
\Psi U^\dagger\Big)} {\int dU\,\e^{N
\tr{}(\Phi U \Psi U^\dagger)}}
=\sum_{n=1}^\infty  F_{n}
\ntr{}\left(t^a\Phi^{n}\right)
\label{Lambda}
\ee
where $\Phi$ and $\Psi$ play
the role of external fields
and $t^a$ ($a=1,\ldots,N^2$--$1$) stand for the generators of the $SU(N)$.
\eq{Lambda} holds~\cite{Mig92a,Mig92d} at $N=\infty$.
The choice of $F_0$ which is not determined by \eq{Lambda}
is a matter of convenience~\cite{Mak93}.

Taking the $1/\l$
term of the expansion of \eq{main} in $\l$ and using Eqs.~\rf{defL}
and \rf{Lambda}, one arrives at the equation for $W(\nu)$
which coincides with \eq{lleq} for the Hermitean one-matrix model
where $V^\p$ is substituted by
\be
\tilde{V}^\p(\l) = \VVp(\l) - F(\l) \,.
\label{tildeVp}
\ee
The potential $\tilde{V}(\om)$
is, generally speaking, non-polynomial and has singularities on the
complex plane outside of the cut (or cuts) of $W(\om)$.

It is worth mentioning that \eq{main} coincides with the loop equation
for the Hermitean two-matrix model~\cite{2mamo}.
This is because at $d=1/2$,
which is associated with the Hermitean two-matrix model, the last
term on the r.h.s.\  of \eq{defL} disappears and
one gets just ${\cal V}(\om)=V(\om)$.

\subsection{The master field equation}

To analyze the model~\rf{spartition},
let us consider the Hermitean two-matrix model with the potential
\be
{\cal  V}(\Phi)=\sum_{m=1}^\infty \frac{g_{m}}{m} \Phi^{m} \,.
\label{def L}
\ee
The solution for $W(\l)$ versus ${\cal  V}(\l)$ is determined by
the equation~\cite{DMS93}
\be
\sum_{m\geq1} g_{m}G_{m-1}(\l) = \l W(\l) -1
\label{smart}
\ee
which is just the $1/\nu$ term of the expansion of \eq{main}
in $1/\nu$.

The functions $G_n(\l)$ are expressed via $W(\l)$ using the recurrence
relation
\be
G_{n+1}(\l)=\ci \frac{\VVp(\om)}{\l-\om}\, G_n(\om) - W(\l)\, G_n(\l)~,
{}~~~~~~G_0(\l)=W(\l)
\label{recurrent}
\ee
which is obtained expanding \eq{main} in $1/\l$.
If ${\cal V}(\l)$ is a polynomial of degree $J$,
\eq{smart} contains $W(\l)$ up to degree $J$ and the solution is
algebraic~\cite{2mamo}.

As is proven in Ref.~\cite{DMS93}:
\begin{itemize}
\item[i)] \vspace{-8pt}
 Equations which appear from
the next terms of the $1/\nu$-expansion of
\eq{main} are automatically
satisfied as a consequence of Eqs.~\rf{smart} and \rf{recurrent}.
\item[ii)] \vspace{-8pt}
$G(\nu,\l)$ is symmetric in $\nu$ and $\l$
for any solution of \eq{smart}.
The symmetry requirement can be used directly to determine $W(\l)$
alternatively to \eq{smart}.
\end{itemize}
\vspace{-7pt}

Since the approach based on \eq{main} is equivalent~\cite{DMS93} to that
of Ref.~\cite{Mig92a},
$G(\nu,\l)$ can be expressed via $W(\l)$ as follows~\cite{Mig92a,Gro92,Bou93}
\be
G(\nu,\l) = 1- \exp{\left\lbrace
\mp \ci \frac{1}{\nu-\om}\log{(\l-r_\pm(\om)})\right\rbrace}
\label{RGsolution}
\ee
where
\be
r_\pm(\l)= \frac {\VVp(\l)+F(\l)}{2} \pm i\pi\rho(\l) =
\left\{
\begin{array}{l}
{\cal V}^\prime (\l)-W(\l) \\ F(\l)+W(\l)
\end{array}
\right.\,.
\label{r}
\ee

The condition for the r.h.s.\ of \eq{RGsolution} to be  symmetric in
$\nu$ and $\l$ is~\cite{Bou93}
\be
r_\pm(r_\mp(\l))=\l \,.
\label{m}
\ee
This relation in $d=1$
was advocated in Ref.~\cite{Mat93} studying the large-$N$ asymptotics of the
integral over the unitary group in~\rf{spartition} (the Itzykson--Zuber
integral).  Since $G(\nu,\l)$ is symmetric in $\nu$ and $\l$, the master
field equation~\cite{Mig92a} \be W(\l) = \pm \ci \log{(\l-r_\pm(\om)})\,,
\label{MFE}
\ee
obtained as the $1/\nu$ term of \eq{RGsolution},
will be satisfied as a consequence of \eq{m}
which guarantees the symmetry.

All three equations~\rf{smart}, \rf{m} and \rf{MFE} seem to be
equivalent. It is a matter of practical
convenience which equation to solve.
The only known explicit solutions exist for the quadratic
potential~\cite{Gro92}
and the logarithmic potential~\cite{Mak93}.

\subsection{Explicit solution for logarithmic potential}

Let us choose the following potential of the Hermitean two-matrix model
\be
{\cal V}(\Phi) =  -(ab+c) \log{(b-\Phi)}  - a \Phi =
\sum_{m=2}^\infty \frac{ab+c}{m\,b^{m}} \Phi^m + \frac cb \Phi
\label{ttV}
\ee
where $a$ and $b$ are real.
In the one-matrix case this potential is associated with the
Penner model~\cite{Pen87}.
The quadratic potential is recovered
in the limit
\be
a,b\ra\infty~,~~~~~~\frac ab  \sim1~,~~~~~~c\sim1
\hbox{ \ \ \ \ (quadratic potential)}~.
\label{limit}
\ee

The solution to \eq{main} for $G(\nu,\l)$ versus $W(\l)$
with ${\cal V}(\om)$ given by \eq{ttV} is
\be
G(\nu,\l) = \frac{W(\nu)-
\frac{(a+\l) W(\l) -1}{b-\nu}}{\l+W(\nu)-\frac{a\nu+c}{b-\nu}} \,.
\label{penG}
\ee
$W(\l)$ is determined by Eqs.~\rf{smart}, \rf{recurrent} which reduce to the
quadratic equation for $W(\nu)$ of the form of \eq{lleq} for the
Hermitean one-matrix model with the logarithmic potential
\be
\tilde{V}(\Phi)= -(ab+c)\log{(b-\Phi)}+ (ab+c+1) \log{(a+\Phi)} -(a+b)\Phi \,.
\label{tV}
\ee
$G(\nu,\l)$ given by \eq{penG} is indeed symmetric in
$\nu$ and $\l$ providing \eq{lleq} with $V=\tilde V$ is satisfied.

Since $\tilde{V}'(\l)$ is known, the function $F(\l)$ can be determined
from \eq{tildeVp} to be
\be
F(\l) = \frac{b\l-c-1}{a+\l} \,.
\label{penLambda}
\ee
Now \eq{defL} determines the potential
\be
V(\Phi)= -(ab+c) \log{(b-\Phi)}- (2d-1) (ab+c+1) \log{(a+\Phi)}
+[(2d-1)b-a] \Phi \,.
\label{V}
\ee
This formula recovers at $d=1/2$ the potential~\rf{ttV} of the two-matrix
model and at $d=0$ the potential~\rf{tV} of the associated one-matrix
model.

In the Gaussian limit~\rf{limit} all the formulas recover the ones for the
solution found by Gross~\cite{Gro92}.
For this reason the one-cut
solution~\rf{one-cut}, \rf{xandy} with the potential~\rf{tV}
is always realized for $W(\l)$ if $a$ and $b$ are
large enough for the points $b$ and $-a$ to lie outside of the cut.

While the above solution for the logarithmic potential was originally
obtained~\cite{Mak93} solving \eq{main}, it is easy to show that it satisfies
\eq{m} and, therefore, the master field equation~\rf{MFE}. Let us notice
for this purpose that $r_\pm(\l)$ given by \eq{r} with our solution for
$\VVp(\l)$, $F(\l)$ and $W(\l)$ satisfy
the following equation
\be
{\cal D}(r_\pm(\l),\l)={\cal D}(\l,r_\mp(\l))=0
\label{mm}
\ee
with
\bea
{\cal D}(\nu,\l)=-\l^2\nu^2+(b-a)\l\nu(\l+\nu)
+ab(\l^2+\nu^2)-
\l\nu(a^2+b^2+2c+1) \non +(\l+\nu)(b-ac+bc)-c^2-b^2
+(ab+c)[(a+b)W(b)-1]\,,
\label{CCC}
\eea
where the constant $W(b)$ depends on the type of the solution of \eq{lleq}
(one-cut or more-than-one-cut solutions).
Since this ${\cal D}(\nu,\l)$ is symmetric in $\nu$ and $\l$,
\eq{mm} implies~\cite{Mat93} that \eq{m} is satisfied.

According to Ref.~\cite{DMS93},
the pair correlator of $U$ and $U^\dagger$ for the potential~\rf{V}
can be calculated
taking the discontinuity of~\rf{penG} both in $\nu$ and $\l$
across the cut (cuts). For the solution~\rf{penG} one
gets~\cite{Mak93}
\be
C(\nu,\l) \equiv \frac{1}{\pi^2 \rho(\nu) \rho(\l)}\,\hbox{Disc}_\l\,
\Di G(\nu,\l) =
\frac{(a+\nu)(a+\l)}{{\cal D}(\nu,\l)}~,~~~~~~~~~\nu,\l\in \hbox{cut}
\label{CC}
\ee
 with ${\cal D}(\nu,\l)$ given by \eq{CCC}.
In the Gaussian
limit~\rf{limit} when $W(b)\ra 1/b$, one recovers the result~\cite{DMS93}
for $C(\nu,\l)$ in the case of the quadratic potential.

Since $\rho(\nu)$ is real for $\nu \in$ cut, the roots of the denominator,
$r_\pm(\nu)$, are complex so that $C(\nu,\l)$ has no singularities at
the cut. The arguments of Ref.~\cite{DKSW93} suggest, therefore, that the
Kazakov--Migdal model with the logarithmic potential~\rf{V} always remains in
the phase with infinite string tension.

The potential~\rf{V} of the $d$-dimensional matrix model admits
the ``naive'' continuum limit when
\be
\Phi=\eps^{\frac d2 -1}\phi~,~~~~~~~~V(\Phi)=\eps^{d}v(\phi)
\label{canonical}
\ee
with $\phi$ and $v(\phi)$ being finite as $\eps\ra0$.
This continuum limit is reached providing
\be
a=b \sim \eps^{\frac d2 - 2}~~~~~~~~\hbox{as}~~~\eps \ra 0
\label{bcubic}
\ee
and results in
a  $d$-dimensional continuum quartic action~\cite{Mak93}.
The procedure of taking the ``naive'' continuum limit
works for $d<4$ where $b$ given by \eq{bcubic} is divergent.
This is precisely
where the scalar theory with the quartic interaction is renormalizable.
Thus, one can look at the logarithmic potential~\rf{V} as at a latticization
of the quartic one for $d<4$.

\subsection{Fermionic matrix models}

The correlators of arbitrary powers of $\bar{\Psi}_{x}\Psi_{x}$
at the same site $x$ are determined for the fermionic
matrix model~\rf{fpartition} by
\be
W_0(\nu)=
\left\langle \ntr{}
\Big( \frac{\nu}{\nu^2-\bar{\Psi}_{x}\Psi_{x}} \Big)
 \right\rangle\,.
\label{defE}
\ee

The analogue of \eq{leq} for the fermionic one-matrix model
(defined by~\rf{fpartition} with $d=0$) reads~\cite{MZ93}
\be
\int_{C_{1}}\frac{d\om}{4\pi i}
\frac{V^\p_F(\om)}{\l-\om}W(\om)=W^{2}(\l)-
\frac{2}{\l}W(\l)+\frac{1}{N^{2}}\frac{\delta}{\delta V_F(\l)}W(\l).
\label{1loopeq}
\ee
This equation is identical to \eq{leq} for the Hermitean one-matrix
model with the logarithmic potential
\be
{V}(\Phi) = V_F(\Phi) +2\log{\Phi}
\label{V+log}
\ee
and $\Phi=\bar{\Psi}\Psi$. These two models are equivalent
to all orders of the $1/N$-expansion. However, the genus expansion has now
alternating signs and is convergent contrary to Ref.~\cite{ds}.

Let us define for the fermionic matrix model~\rf{fpartition}
on a $d$-dimensional lattice the odd-odd
and even-even one-link correlators:
\bea
{G}(\nu,\l) &= &P_\mu^\pm
\left\langle \ntr{}
\Big(  \Psi_x \frac{1}{\nu^2-\bar{\Psi}_{x}\Psi_{x}}U_\mu(x)
\frac{1}{\l^2-\bar{\Psi}_{x+\mu}\Psi_{x+\mu}} \bar{\Psi}_{x+\mu}
U^\dagger_\mu(x) \Big)
 \right\rangle \,, \non
W(\nu,\l)&= &
\left\langle \ntr{}
\Big( \frac{\nu}{\nu^2-\bar{\Psi}_{x}\Psi_{x}}U_\mu(x)
\frac{\l}{\l^2-\bar{\Psi}_{x+\mu}\Psi_{x+\mu}} U^\dagger_\mu(x) \Big)
 \right\rangle \,,
\label{defH}
\eea
where $+(-)$ are associated with the positive (negative) direction $\mu$.

The fermionic analogue of \eq{main} consists of two equations~\cite{MZ93}
\bea
 \int_{C_1} \frac{d \om}{4\pi i}
\frac{\VVp(\om)}{\nu - \om}G(\om, \l) &= &
W_0(\nu) G(\nu, \l) +  \Big[\l W(\nu, \l) - W_0(\nu) \Big] \,,
\label{2main1} \non
 \int_{C_1} \frac{d \om}{4\pi i}
\frac{\VVp(\om)}{\nu - \om}W(\om, \l) & = & W_0(\nu) W(\nu, \l)
- 2 \frac{1}{\nu} W (\nu, \l)
 - \l  G(\nu, \l)  \,.
\label{2main2}
\eea

The solution to Eq.~\rf{2main2} for the quadratic potential
$V_F(\om)=m\,\om^2$ is
\be
W_{0}(\l)=\frac{1}{2}
\left[\, \mu\l+\frac{2}{\l}-\frac{1}{\l}
\sqrt{\mu^{2}\l^{4}+4}\, \right]
\label{DGauss}
\ee
 with
\be
\mu=\frac{(D-1)m+D\sqrt{m^{2}+4 (2D-1)}}{(2D-1)}~.
\label{mu}
\ee
This solution agrees~\cite{KhM92b} with the result~\cite{KMNP81} for
lattice QCD with fundamental fermions at vanishing plaquette term.

\newsection{Conclusions}

There is no problems to calculate higher orders of genus expansion in the
Hermitean one-matrix model using the iterative method which is described in
Sect.~2.  It would be interesting to perform analogous calculations for the
potential~\rf{V+log} which is equivalent to the fermionic model whose
genus expansion is expected to be convergent since
the integral over the Grassmann variables
in~\rf{fpartition} converges.

The method of solving the Kazakov--Migdal model which is described in Sect.~3
reduces it at large-$N$ to the Hermitean two-matrix model. This goes along
with Ref.~\cite{DKK93} where the conformal field theories in $d\leq1$ are
obtained from the two-matrix model. For $d>1$ the potential ${\cal V}$ of
the two-matrix model should be, presumably, non-polynomial rather than
polynomial as for $d\leq1$.

The approach of Sect.~3 works, however, only when singularities of ${\cal V}$
lie outside of the cut. This is not the case for the solution of the
 Riemann--Hilbert equations found by Migdal~\cite{Mig92a} which exhibits a
non-trivial critical behavior. While this solution is not associated with
induced QCD, it would be interesting to find out what physical system it
corresponds to.

The discussed solution of the $d>1$ models are associated with the strong
coupling phase, \ie the phase with infinite string tension. An open question is
whether the described approach can be extended to the phase with area law where
QCD is induced.

\subsection*{Acknowledgements}
I thank the organizers for the kind hospitality at Kyffhaeuser.


\end{document}